\documentclass[aps,prl,reprint,superscriptaddress,amsmath,amssymb,letterpaper]{revtex4-1}

\usepackage{graphicx,subfigure,float,dcolumn}
\usepackage{bm,color,xcolor,calc}
\usepackage{tikz,tikz-3dplot,pgfplots,pgfplotstable}
\usetikzlibrary{arrows,arrows.meta,positioning,shapes,calc,3d,fadings,matrix,decorations.pathreplacing}
\usepackage{natbib}
\usepackage{pdfpages}
\makeatletter
\AtBeginDocument{\let\LS@rot\@undefined}
\makeatother
\begin{document}
\title{Discovering topological surface states of Dirac points}%

\author{Hengbin Cheng}
\affiliation{Institute of Physics, Chinese Academy of Sciences/Beijing National Laboratory for Condensed Matter Physics, Beijing 100190, China}
\affiliation{School of Physical Sciences, University of Chinese Academy of Sciences, Beijing 100049,China}

\author{Yixin Sha}
\affiliation{School of Electronics Engineering and Computer Science, Peking University, Beijing 100871, China}

\author{Rongjuan Liu}
\affiliation{Institute of Physics, Chinese Academy of Sciences/Beijing National Laboratory for Condensed Matter Physics, Beijing 100190, China}

\author{Chen Fang}
\affiliation{Institute of Physics, Chinese Academy of Sciences/Beijing National Laboratory for Condensed Matter Physics, Beijing 100190, China}
\affiliation{Songshan Lake Materials Laboratory, Dongguan, Guangdong 523808, China.}

\author{Ling Lu}
\email{linglu@iphy.ac.cn}
\affiliation{Institute of Physics, Chinese Academy of Sciences/Beijing National Laboratory for Condensed Matter Physics, Beijing 100190, China}
\affiliation{Songshan Lake Materials Laboratory, Dongguan, Guangdong 523808, China.}
\begin{abstract}
Dirac materials, unlike the Weyl materials, have not been found in experiments to support intrinsic topological surface states, as the surface arcs in existing systems are unstable against symmetry-preserving perturbations. Utilizing the proposed glide and time-reversal symmetries, we theoretically design and experimentally verify an acoustic crystal of two frequency-isolated three-dimensional Dirac points with $Z_2$ monopole charges and four gapless helicoid surface states.
\end{abstract}
\date{\today}
\maketitle
A three-dimensional~(3D) Dirac point~\cite{armitage2018weyl} disperses the same way as the solutions to the massless Dirac equation at the vicinity of the four-fold linear point degeneracy.
Playing a central role in 3D band topology, Dirac points can, upon symmetry breaking, transition into Weyl points, line nodes or topological bandgaps with gapless surface states.
Although 3D Dirac points have been experimentally discovered in electron~\cite{liu2014discovery,neupane2014observation,liu2014stable,s2014experimental,xu2015observation,yi2015evidence}, magnon~\cite{yao2018topological,bao2018discovery} and photonic~\cite{guo2019observation} systems along with a variety of other theoretical proposals~\cite{young2012dirac,wang2012dirac, wang2013three,steinberg2014bulk,wang2016three,tang2016dirac,li2017topological,slobozhanyuk2017three,guo2017three,wang2017type,le2018dirac,zhang2019multiple,yang2019realization}, none of the surface states are topological. Specifically, there have been no robust gapless surface bands associated with the bulk Dirac points~\cite{kargarian2016are,kargarian2018deformation,wu2019fragility}. 

\begin{figure*}[t!]
    \includegraphics[width=16.5cm,height=12.0cm]{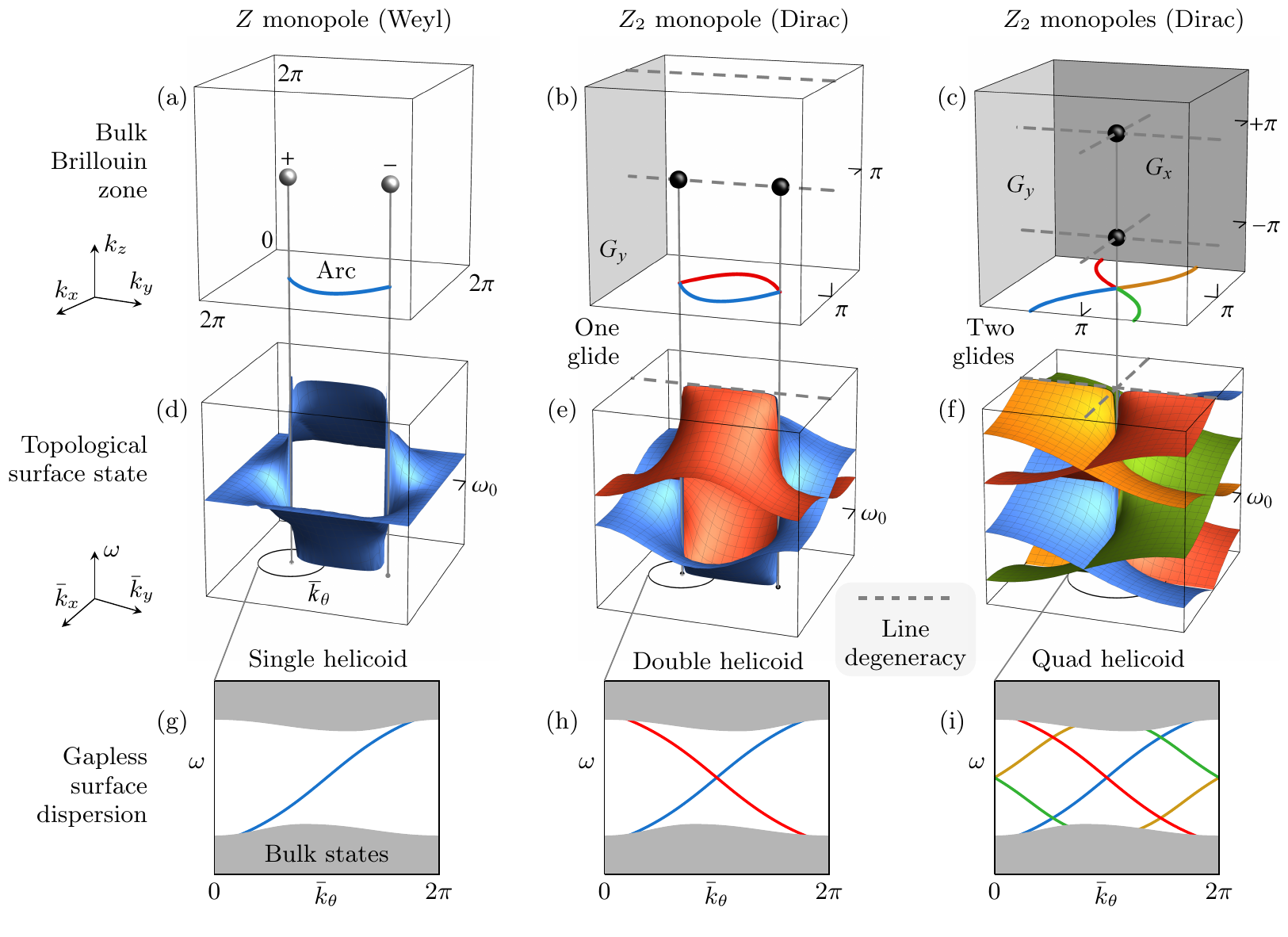} 
    \caption{Bulk monopoles and helicoid surfaces.
    (a), (b), (c) are the 3D BZ containing two $Z$ or $Z_2$ monopoles. (a), (b) can be the BZs of cubic lattices, while (c) is the BZ of the BCC lattice in our acoustic design. The arcs, at the bottom plane, are the iso-frequency contours of the helicoid surface states at frequency $\omega_0$ labeled in (d), (e), (f). The glide planes are fill in gray and the $G_i\mathcal{T}$ protected line degeneracies are shown in dashed gray lines.
    (g), (h), (i) illustrates the gapless surface dispersions, around the projected monopoles, along a circular path of $\bar{k}_\theta$ in the surface BZ.
    The $Z$ monopoles can be Weyl points, nodal lines or surfaces. The $Z_2$ monpoles can be Dirac points, $Z_2$ nodal lines or Weyl dipoles.}
	\label{fig:helicoid}
\end{figure*}

The current lack of topological surface states for Dirac points can be understood through the anticrossing of two Weyl surface states. Illustrated in Fig.~\ref{fig:helicoid}, the topological surface dispersion of a Weyl crystal is a doubly-periodic helicoid sheet whose singularities locates at the projection of the bulk Weyl points.
The chirality of the helicoid around each Weyl point equals the sign of its Chern number.
Since a Dirac point is composed of two Weyl points of opposite Chern numbers, the Dirac surface state should be composed of two helicoids of opposite chiralities. Two opposite helicoid surfaces generally cross each other along a line of momenta and anticross~(hybridize with each other), resulting in gapped surface bands which are topologically trivial.
The only exception was theoretically proposed in Ref.~\cite{fang2016topological}, in which the glide symmetries combined with the time-reversal~($\mathcal{T}$) can stabilize a degenerate line and protect the crossing of the helicoids. As illustrated in Fig.~\ref{fig:helicoid}, one glide can protect double helicoids and two glides can protect quad-helicoid surfaces states.

In this work, we present an acoustic band structure with two ideal $Z_2$ Dirac points protected by glide reflections.
The acoustic crystal is 3D-printed and the measured surface dispersions exhibit quad-helicoid surface sheets.

\paragraph{Ideal acoustic Dirac points}
The cubic cell of the acoustic crystal, in Fig.~2(a), consists of thick rods and thin sticks, belonging to space group $Ia\bar{3}$~(No.~206) of the body-centered-cubic~(BCC) lattice. The four thick rods of radius $0.15a$ point at the directions of the BCC lattice vectors, where $a$ is the lattice constant of the cubic cell. These disconnected rods form the BPI~(blue phase I of liquid crystal) photonic crystal in Ref.~\cite{lu2016symmetry,lu2015generalized}. We add the thin sticks to connect all rods and mechanically support the whole structure. The sticks are too thin, $0.025a$ in radius, to change the Dirac acoustic bands, as compared in Supplementary Materials.
The background material is air and the interfaces are treated as sound hard-wall boundaries in numerical simulations.

\begin{figure*}[t!]
	\includegraphics[width=16.8cm,height=9.0cm]{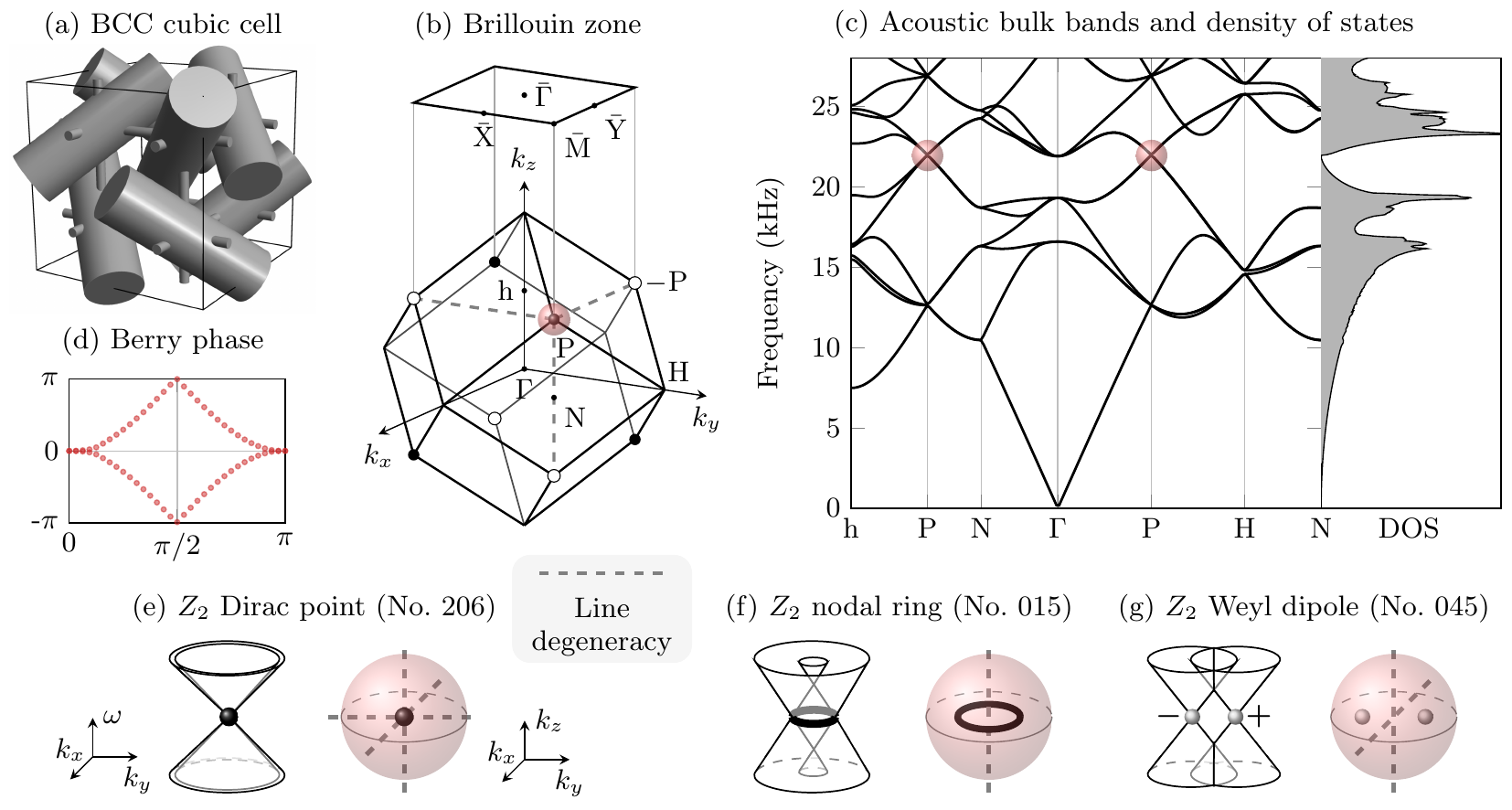}
	\caption{Ideal acoustic Dirac points with $Z_2$ charges.
	(a)~One cubic cell of the acoustic crystal of BCC lattice. (b)~BCC BZ and the $[001]$ surface BZ. The two $\pm P$ points project to the same point $\rm \bar{M}$ in the surface BZ. (c)~Acoustic bulk band structure with the Dirac points crossing between the 5-6th and 7-8th bands at the $P$ point. The DOS vanished at the frequency of ideal Dirac points. (d)~Calculated non-Abelian Berry phases on the surface of the red sphere enclosing the Dirac point. Over the polar angle, the nontrivial winding of Berry phases implies topological invariant of $Z_2=1$. (e, f, g)~Three types of four-bands nodal structures that carry $Z_2$ monopole charges. The dashed gray lines are the line degeneracies due to $G_i\mathcal{T}$, and the red spherical surfaces enclose the $Z_2$ monopoles.}
		\label{fig:theory}
\end{figure*}

There are two Dirac points locate at the $\pm P$ points of the BCC Brillouin zone~[BZ, Fig.~2(b)], where four bands~(from the fifth to the eighth in ascending energy order) meet, as shown in Fig.~\ref{fig:theory}(c).
The density of states~(DOS)~\cite{liu2018generalized} vanishes at the Dirac frequency and grows quadratically away from it, as expected for linear dispersion relations.

The local Hamiltonian of this Dirac point is $H(\boldsymbol{k})\sim(\begin{smallmatrix}\boldsymbol{k}\cdot\boldsymbol{\sigma} & 0\\0 & -\boldsymbol{k}\cdot\boldsymbol{\sigma}\end{smallmatrix})$ , determined by the $k\cdot p$ analysis detailed in the Supplementary Materials.
This four-fold degeneracy is joined, due to the anti-unitary parity-time symmetry~($\mathcal{PT}$), by two conjugated 2D representations of the little group. The little group of No.~206 at $P$ also has a four fold representation which is the generalized Dirac point discussed in Ref.~\cite{lu2015generalized,lu2016symmetry}. The Dirac point has identical group velocities while the generalized one does not.

Each Dirac point is stabilized by the products of $\mathcal{T}$ and the three non-commuting glides~($G_x=\{M_x|(a/2)\hat{x}+(a/2)\hat{y} \},G_y=\{M_y|(a/2)\hat{x}\} ,G_z=\{M_z|(a/2)\hat{y} \}$), denoted as $G_i\mathcal{T}~(i=x,y,z)$.
$M_i$ are the mirror operations and the inversion $\mathcal{P}=G_xG_yG_z$.
Each $G_i\mathcal{T}$ symmetry enforces a line degeneracy at the zone boundary, represented by the dashed lines in all figures consistently. The three degeneracy lines intersect at the $P$ point, shown in Fig.~\ref{fig:theory}(b), (e).
We note that the $P$ point have neither $\mathcal{T}$ nor $G_i$ symmetries by themselves.

\paragraph{$Z_2$ monopoles}
Dirac point is the symmetry-protected $Z_2$ monopole in the 3D momentum space~\cite{morimoto2014weyl,yang2014classification,yang2015topological}. However, the $Z_2$ symmetries in the these systems~($\mathcal{PT}$ for example) cannot protect any line degeneracies on the surface, disallowing the gapless connectivity between two helicoids of opposite chiralities.
In contrast, the $Z_2$ invariant of our acoustic Dirac point is protected by $G_i\mathcal{T}$~\cite{fang2016topological}, leading to the nontrivial band topology not only in the bulk but also on the surface.

In Fig.~\ref{fig:theory}(d), we calculate the non-Abelian Berry phase~\cite{yu2011equivalent} of the lower two bands~(5th and 6th) on a sphere enclosing the Dirac point. The gapless spectra indicate the nontrivial monopole charge of $Z_2=1$. 
Since this $Z_2$ charge can be protected by either one of the three $G_i\mathcal{T}$~\cite{fang2016topological}, we can break the other two or one $G_i\mathcal{T}$ to get the $Z_2$ nodal ring~\cite{fang2015topological,li2017dirac,bzdusek2017robust,song2018diagnosis,ahn2018band} and Weyl dipoles~\cite{morimoto2014weyl,fang2016topological}. These symmetry-breaking cases are illustrated in Fig.~\ref{fig:theory}~(e),~(f),~(g)and discussed in detail in the Supplementary Materials.
\paragraph{Quad-helicoid and Jacobi elliptic function}
We project the two Dirac points onto the (001) surface, corresponding to the case in Fig.~\ref{fig:helicoid}~(c),~(f),~(i).
The plane group of this surface is $p2gg$, on which the two degenerate lines due to $G_x\mathcal{T}$ and $G_y\mathcal{T}$ are presented.
These two line degeneracies, outlining the whole boundaries of the surface BZ, protect all the crossings among the four helicoid surface sheets.
The iso-frequency contour, in Fig.~\ref{fig:helicoid}(c), are four branches originating from the projected Dirac points. The four branches are connected across the zone boundary forming two non-contractable loops.
\begin{figure*}[!ht]
	\includegraphics[width=17.7cm,height=11.8cm]{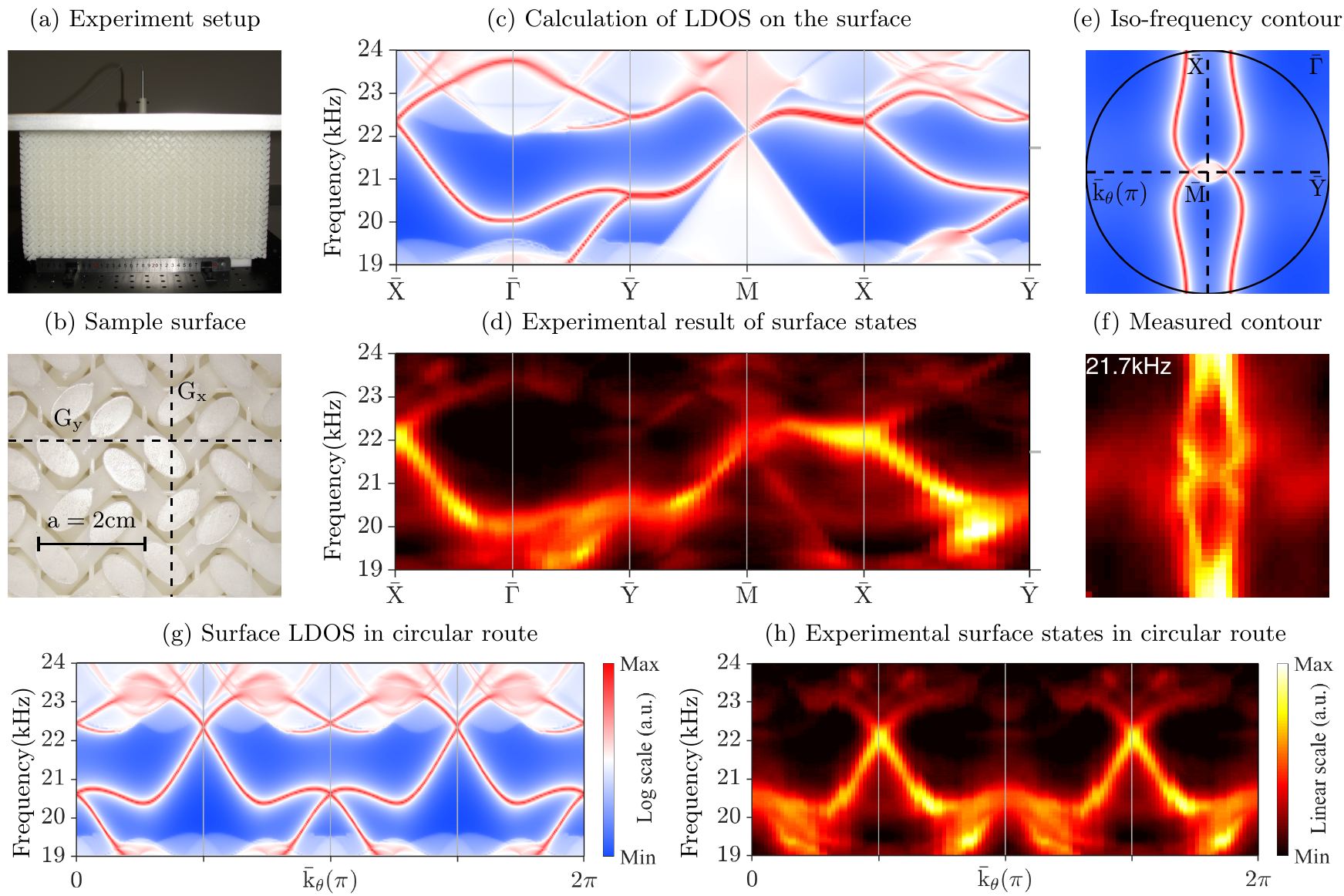}
	\caption{Experiment observation of the quad-helicoid surface states by FTFS. (a)~Photograph of the 3D-printed sample on the measurement setup. (b)~Photograph of the sample surface.
	(c), (d)~Numerical and experimental results of the surface states on the high-symmetry momentum lines. (e), (f)~Numerical and experimental iso-frequency contour of the quad-helicoid surfaces showing four branches at 21.7kHz. (g), (h)~Numerical and experimental results of the surface states in a circular route around the projected bulk Dirac points.}
	\label{fig:exp}
\end{figure*}

If we parametrize the 2D surface BZ as a complex plane~($z\propto k_x+ik_y$)~\cite{fang2016topological}, the helicoid surface bands can be expressed as~(are topologically equivalent to) the double-periodic elliptic functions~\cite{zhang2018double,yang2018ideal}.
The four helicoids in Fig.~\ref{fig:helicoid}(f) are plotted using the Jacobi elliptic functions $cn(z,\frac{1}{2})$.
The Jacobi function has two zeros and two poles in one period, each locating at the center of each quadrant. These four singularities represent two Weyl dipoles and all project to the same point in the surface BZ. So we construct the quad-helicoid surfaces by stacking the four quadrants of the Jacobi function and align the central singularities.
The mapping of one quadrant is $\omega(k_x, k_y) \sim \rm{Im}\mit(\log[cn(z_i(k_x,k_y), \frac{1}{2})])$, where $z_1(k_x,k_y)=\frac{\rm K(\frac{1}{2})}{\pi}(\frac{k_x-k_y}{2}+i\frac{k_x+k_y}{2})$ and $\rm K$ is the complete elliptic integrals of the first kind. The rest three quadrants are obtained by the translations of $z_2(k_x,k_y)=z_1(k_x+2\pi,k_y)$, $z_3(k_x,k_y)=z_1(k_x,k_y+2\pi)$, and $z_4(k_x,k_y)=z_1(k_x+2\pi,k_y+2\pi)$.

\paragraph{Experiments}
A photograph of the Dirac acoustic crystal is shown in Fig.~\ref{fig:exp}(a), 3D-printed by the stereo lithography method using photocurable resin. The lattice constant is $a=20$mm and the fabrication error is $\pm$ 0.1mm. The total size of the sample is 413.0mm$\times$413.0mm$\times$222.8mm containing 20$\times$20$\times$11 cubic cells.

The surface states are measured through the Fourier-transformed field scan~(FTFS). Similar approaches have been used to study other topological acoustic crystals~\cite{li2018weyl,ge2018experimental,yang2019topological,peri2019axial,xie2019experimental}. A pressure-field microphone~(diameter of 3.5mm, B\&K-4138-A-015) is used as the scanning probe~(receiver). The microphone is embedded in an aluminum alloy plate which works as a hard wall boundary on the top surface of the sample, as show in Fig.~\ref{fig:exp}(a). The acoustic source is a broadband earphone, having frequency response up to 40kHz and a diameter of 5.5mm, fixed at the corner of the sample close to the plate. 
The amplitude and phase of the pressure field are collected by the data acquisition module B\&K-3160-A-042. A broadband signal is generated from the module and split into two channels, one to driven the earphone and the other as a time reference for the receiver. 
The frequency spectrum is averaged 100 times for each point scan and is normalized by the signal from the source.

The field scan is performed by moving the sample stage in three directions. During the collection of each data point, the sample is pressed towards the top plate to ensure the absence of air gaps.
The scanning step is 5mm and the scanning range is 400mm in both $x$ and $y$ directions. Through 2D Fourier transforms, we obtain the spectral weight of the surface states in the momentum range of (-2, 2)$\frac{2\pi}{a}$. 
Similar to the processing technique used in Ref.~\cite{yan2018experimental}, we patch the data of spatial scans to double the momentum resolution in the reciprocal space.
In the $x$ direction, we stitch two scanning fields of equivalent source positions. In the $y$ direction, we rotate the data due to the $C_2$ symmetry on the surface.

The FTFS results are shown in three plots in Fig.~\ref{fig:exp}~(d)~(f)~(h). The corresponding numerical results of the local density of states~(LDOS)\cite{Sancho_1985,wu2017} at the measurement interface are shown in Fig.~\ref{fig:exp}~(c)~(e)~(g) respectively. The detailed algorithm of the surface LDOS will be presented in a separate paper.
The agreement between experiments and numerics are visually obviously for the gapless surface dispersions.

\paragraph{Discussion}
We experimentally observed the first example of topological surface states associated with the 3D Dirac points. The line-degeneracy due to glide and $\mathcal{T}$ symmetry is the key for stabilizing the gapless connection of helicoid sheets of opposite chiralities. Similarly, the other nonsymmorphic symmetry~(screw rotation) and $\mathcal{T}$ could also protect such line degeneracies when a domain wall is constructed to preserve the screw axis on the surface. 
It will also be interesting to explore the material realization of the double-helicoid surface states shown in Fig.~\ref{fig:helicoid}(e), as well as helicoid surface states of Dirac semimetals~\cite{fang2016topological}.

This work establishes an ideal 3D Dirac material for consequent studies. For example, symmetry breakings of Dirac points can generate a variety of topological phenomena, such as the charged~($Z$ or $Z_2$) nodal lines, nodal surfaces~\cite{xiao2017topologically}, Weyl dipoles as well as a gapped bulk state supporting gapless surface~\cite{lu2016symmetry} or chiral hinge modes~\cite{yue2019symmetry,kim2019glide}.

\paragraph{Acknowledgements}
We thank Timothy Hsieh and Liang Fu for the previous discussions on the $k\cdot p$ model.
We are supported by the National key R\&D Program of China under Grant No. 2017YFA0303800, 2016YFA0302400, 2016YFA0300600, by NSFC under Project No. 11974415, 11721404, 11674370, and by Chinese Academy of Sciences under grant number XXH13506-202.

\bibliography{references.bib}
\bibliographystyle{unsrt}



\section*{Supplementary material}

# Discovering topological surface states of Dirac points:
# Supplementary material

Hengbin Cheng,[1,2] Yixin Sha,[3] Rongjuan Liu,[1] Chen Fang,[1] and Ling Lu[1,4]

[1]*Institute of Physics, Chinese Academy of Sciences/Beijing National
Laboratory for Condensed Matter Physics, Beijing 100190, China*
[2]*School of Physical Sciences, University of Chinese Academy of Sciences, Beijing 100049, China*
[3]*School of Electronics Engineering and Computer Science, Peking University, Beijing 100871, China*
[4]*Songshan Lake Materials Laboratory, Dongguan, Guangdong 523808, China.*

## I. UNIT CELL DESIGN

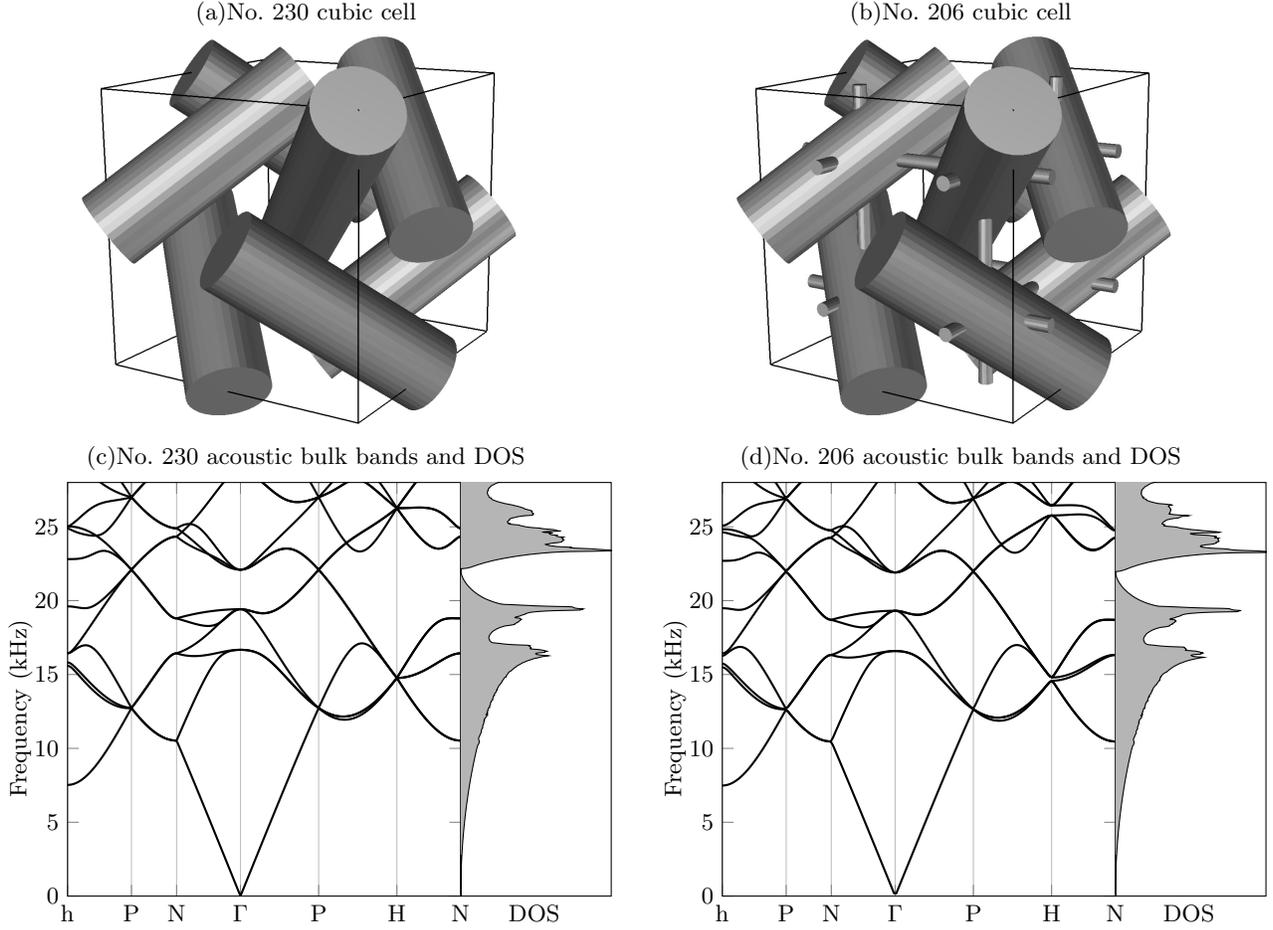

FIG. S1: Design of ideal acoustic Dirac point without and with the thin sticks. (a), (c) BPI acoustic crystal of space group No. 230. (b), (d) Thin-stick joined BPI acoustic crystal of space group No. 206.

For sample stability, we add the 12 thin sticks of radius $0.025a$ (1mm in experiment) to connect all the BPI rods. Fig. S1(a), (b) are the unit-cell structures without and with thin sticks, they belongs to space group No. 230 and No. 206 respectively. The thin sticks preserve the three glides that protect the ideal Dirac points. The corresponding acoustic bulk bands are plotted in Fig. S1(c), (d), which are almost identical for the frequency range of the Dirac points.

## II.  k · p MODEL

The local dispersion and topology at $P$ point can be described by the effective Hamiltonian of $\boldsymbol{k} \cdot \boldsymbol{p}$ model, written as $H(\boldsymbol{k}) = A\boldsymbol{k} \cdot \boldsymbol{p}$, considering the first non-zero order of $\boldsymbol{k}$. Here, we show the detailed derivation of the effective Hamiltonian of our ideal acoustic Dirac point in space group $Ia\bar{3}d$ (No. 230) and $Ia\bar{3}$ (No. 206).

Considering the space group $Ia\bar{3}d$ (No. 230), we choose four independent symmetry operations (generators) of the little group at momentum point $P$. They are two-fold rotation symmetry $C_{2z} = \{C_2|(0, \frac{1}{2}, 0)\}$ along [001] direction, two-fold rotation symmetry $C_{2y} = \{C_2|(\frac{1}{2}, 0, 0)\}$ along [010] direction, three-fold rotation symmetry $C_3$ along [111] direction and glide reflection symmetry $G = \{M|(-\frac{3}{4}, -\frac{3}{4}, \frac{1}{4})\}$ in $[1\bar{1}0]$ plane. Here $G$ is different from the $G_i$ ($i = x, y, z$) glides in [100], [010], [001] planes. $P$ point has no $G_i$ ($i = x, y, z$) symmetries, but has $G_i\mathcal{T}$ that can be obtained by $G_i\mathcal{T} = C_{2i} \cdot \mathcal{PT}$ ($i = x, y, z$). $\mathcal{PT}$ is the parity-time symmetry preserves in the whole BZ.

$P$ point has $\mathcal{PT}$ but has neither $\mathcal{P}$ nor $\mathcal{T}$. $\mathcal{PT}$ is an anti-unitary operator which can be written as $\mathcal{PT} = U \cdot \mathcal{K}$, where $U$ is a unitary matrix and $\mathcal{K}$ is the complex conjugate. The little group of $P$ point has two conjugated 2D representations, forming a 4D representation under $\mathcal{PT}$. We obtain the irreducible representation (Irrep) matrices of the selected generators from Bilbao crystal server, noted as Bilbao Rep. in Table II. After unitary transformations, we have the representation matrices in real basis (Real Rep.) and in Dirac basis (Dirac Rep.).

| Operators | Bilbao Rep. | Real Rep. | Dirac Rep. |
|---|---|---|---|
| $C_{2z}$ | $\begin{pmatrix} i & 0 & 0 & 0 \\ 0 & -i & 0 & 0 \\ 0 & 0 & i & 0 \\ 0 & 0 & 0 & -i \end{pmatrix}$ | $\begin{pmatrix} 0 & 0 & 0 & 1 \\ 0 & 0 & 1 & 0 \\ 0 & -1 & 0 & 0 \\ -1 & 0 & 0 & 0 \end{pmatrix}$ | $\begin{pmatrix} i & 0 & 0 & 0 \\ 0 & -i & 0 & 0 \\ 0 & 0 & i & 0 \\ 0 & 0 & 0 & -i \end{pmatrix}$ |
| $C_{2y}$ | $\begin{pmatrix} 0 & i & 0 & 0 \\ i & 0 & 0 & 0 \\ 0 & 0 & 0 & i \\ 0 & 0 & i & 0 \end{pmatrix}$ | $\begin{pmatrix} 0 & 0 & -1 & 0 \\ 0 & 0 & 0 & 1 \\ 1 & 0 & 0 & 0 \\ 0 & -1 & 0 & 0 \end{pmatrix}$ | $\begin{pmatrix} 0 & -1 & 0 & 0 \\ 1 & 0 & 0 & 0 \\ 0 & 0 & 0 & -1 \\ 0 & 0 & 1 & 0 \end{pmatrix}$ |
| $C_3$ | $\frac{1}{\sqrt{2}}\begin{pmatrix} e^{i\frac{3}{4}\pi} & e^{i\frac{7}{4}\pi} & 0 & 0 \\ e^{i\frac{5}{4}\pi} & e^{i\frac{5}{4}\pi} & 0 & 0 \\ 0 & 0 & e^{i\frac{3}{4}\pi} & e^{i\frac{7}{4}\pi} \\ 0 & 0 & e^{i\frac{5}{4}\pi} & e^{i\frac{5}{4}\pi} \end{pmatrix}$ | $\frac{1}{2}\begin{pmatrix} -1 & -1 & -1 & -1 \\ 1 & -1 & -1 & 1 \\ 1 & 1 & -1 & -1 \\ 1 & -1 & 1 & -1 \end{pmatrix}$ | $\frac{-1}{2}\begin{pmatrix} 1-i & -1-i & 0 & 0 \\ 1-i & 1+i & 0 & 0 \\ 0 & 0 & 1-i & -1-i \\ 0 & 0 & 1-i & 1+i \end{pmatrix}$ |
| $G$ | $\begin{pmatrix} 0 & 1 & 0 & 0 \\ i & 0 & 0 & 0 \\ 0 & 0 & 0 & -1 \\ 0 & 0 & -i & 0 \end{pmatrix}$ | $\frac{-1}{2}\begin{pmatrix} 0 & 1+i & 0 & 1+i \\ 1+i & 0 & -1-i & 0 \\ 0 & -1-i & 0 & 1+i \\ 1+i & 0 & 1+i & 0 \end{pmatrix}$ | $\begin{pmatrix} 0 & 0 & 0 & -i \\ 0 & 0 & -1 & 0 \\ 0 & -i & 0 & 0 \\ -1 & 0 & 0 & 0 \end{pmatrix}$ |
| $\mathcal{PT}$ | $\Gamma_{22}\mathcal{K}$ | $\mathcal{K}$ (complex conjugate) | $\Gamma_{22}\mathcal{K}$ |
| $H(\boldsymbol{k})$ | $(\sigma_x + w\sigma_y)\otimes$ $(k_z\sigma_z + k_x\sigma_y - k_y\sigma_x)$ | $(-k_x\Gamma_{10} - k_y\Gamma_{33} + k_z\Gamma_{31})$ $+w(-k_x\Gamma_{30} + k_y\Gamma_{13} - k_z\Gamma_{11})$ | $\sqrt{1+w^2}\sigma_z \otimes \boldsymbol{k} \cdot \boldsymbol{\sigma}$ |
| mass | $m\Gamma_{30} + m'(\Gamma_{20} - w\Gamma_{10})$ | $m\Gamma_{20} + m'(\Gamma_{32} - w\Gamma_{12})$ | $m\Gamma_{10} + m'\Gamma_{20}$ |

TABLE I: Representation matrices of selected generators of the little group at $P$ in space group No. 230. The four matrices in the gray cells are obtained from Bilbao website and the rest are worked out by commutation relations.

We need to find the representation matrix for $\mathcal{PT}$ which is not given by Bilbao. We expand the 4D unitary $U$ matrix by the 16 Gamma matrices with complex coefficients and determine the coefficients by the commutation relations in Table II.

| | $\mathcal{PT} = U \cdot \mathcal{K}$ | | $C_{2z}$ | $C_{2y}$ | $G$ | $C_3$ |
|---|---|---|---|---|---|---|
| $\mathcal{PT}$ | $\mathcal{PT}^2 = 1$ | $UU^\dagger = 1$ | $[C_{2z}, \mathcal{PT}] = 0$ | $[C_{2y}, \mathcal{PT}] = 0$ | $G(\mathcal{PT}) = i(\mathcal{PT})G$ | $[C_3, \mathcal{PT}] = 0$ |

TABLE II: Compatibility relations used to determine $\mathcal{PT}$.

The sixteen Gamma matrices are $\Gamma_{ij} = \sigma_i \otimes \sigma_j$, ($i, j = 0, 1, 2, 3$), where $\sigma_0 = \begin{pmatrix} 1 & 0 \\ 0 & 1 \end{pmatrix}, \sigma_1 = \begin{pmatrix} 0 & 1 \\ 1 & 0 \end{pmatrix}, \sigma_2 = \begin{pmatrix} 0 & -i \\ i & 0 \end{pmatrix}, \sigma_3 = \begin{pmatrix} 1 & 0 \\ 0 & -1 \end{pmatrix}$, and $\otimes$ is the Kronecker product.

The Hamiltonian can be written as $H(\boldsymbol{k}) = A(k_x p_x + k_y p_y + k_z p_z)$, where $k_x$, $k_y$, $k_z$ are three real variables (the origin is set at $P$ in below), and $p_x$, $p_y$, $p_z$ are three Hermitian matrices. To find $p_x$, $p_y$, $p_z$, we expand them as

linear combinations of 16 Gamma matrices with real coefficients and determine the coefficient by the commutation relations in Table. III.

For any representation matrix $D(g)$ of symmetry operation $g$ (including $\mathcal{PT}$) in the little group, we have

$$D(g)H(\mathbf{k})D^{-1}(g) = H(g\mathbf{k}) \quad (1)$$

that gives the following commutation ([]=0) and anti-commutation ({}=0) relations, listed in Table. III.

|  | $p_z$ | $p_y$ | $p_x$ |
|---|---|---|---|
| $\mathcal{PT}$ | $[\mathcal{PT}, p_z] = 0$ | $[\mathcal{PT}, p_y] = 0$ | $[\mathcal{PT}, p_x] = 0$ |
| $C_{2z}$ | $[C_{2z}, p_z] = 0$ | $\{C_{2z}, p_y\} = 0$ | $\{C_{2z}, p_x\} = 0$ |
| $C_{2y}$ | $\{C_{2y}, p_z\} = 0$ | $[C_{2y}, p_y] = 0$ | $\{C_{2y}, p_x\} = 0$ |
| $G$ | $[G, p_z] = 0$ | $Gp_y = p_x G$ | |
| $C_3$ | $p_y = C_3^{-1} p_z C_3, \ p_x = C_3 p_z C_3^{-1}$ | | |

TABLE III: Compatibility relations used to determine $H(\mathbf{k})$.

By satisfying all the requirements in Table. III, for Bilbao Rep. we obtain $p_z = \Gamma_{13} + w\Gamma_{23}$, $(w \in \mathbb{R})$. $p_x$ and $p_y$ are obtained from $p_z$ by $C_3$ symmetry. Ignore the constant term $A$, the effective Hamiltonian is given as

$$H_D(\mathbf{k}) \sim (\sigma_x + w\sigma_y) \otimes (k_z \sigma_z + k_x \sigma_y - k_y \sigma_x) \quad (2)$$

After an unitary transformation, we get the representation matrix under Dirac Rep.

$$H_D(\mathbf{k}) \sim \sqrt{1+w^2} \begin{pmatrix} \mathbf{k} \cdot \boldsymbol{\sigma} & 0 \\ 0 & -\mathbf{k} \cdot \boldsymbol{\sigma} \end{pmatrix} \quad (3)$$

where $w \in \mathbb{R}$ is the velocity term. Note that the Dirac Hamiltonian we obtained above satisfies the exact massless Dirac equation in 3D. What listed in the last row of Table. I are the matrix of the Dirac mass terms, in corresponding basis with coefficients $m, m' \in \mathbb{R}$.

For space group No. 206, the only missing generators is $G = \{M|(-\frac{3}{4}, -\frac{3}{4}, \frac{1}{4})\}$ in $[1\bar{1}0]$ plane, and the Dirac Hamiltonian under Bilbao Rep. is

$$H_D(\mathbf{k}) \sim (\sigma_x + w\sigma_y + u\sigma_z) \otimes (k_z \sigma_z - k_x \sigma_x - k_y \sigma_y), \ (w, u \in \mathbb{R}) \quad (4)$$

. After an unitary transformation, we have the representation matrix under Dirac Rep.

$$H_D(\mathbf{k}) \sim \sqrt{1+w^2+u^2} \begin{pmatrix} \mathbf{k} \cdot \boldsymbol{\sigma} & 0 \\ 0 & -\mathbf{k} \cdot \boldsymbol{\sigma} \end{pmatrix} \quad (5)$$

## III. DIRAC POINT UNDER SYMMETRY BREAKINGS

Dirac point lies in the phase transition center of many topological band nodal structures. In Table. IV we list the possible four bands nodal structures that can be achieved from perturbing the Dirac point, and list the corresponding key symmetries that need to be preserved($\sqrt{}$) or broken($\times$).

| | $Z_2$ Dirac point | $Z_2$ nodal ring | $Z_2$ Weyl dipole | gapped Dirac | gapped | Weyl points | nodal ring |
|---|---|---|---|---|---|---|---|
| $Z_2$ | 1 | | | | 0 | | |
| Realization | Fig. 2(a) | Fig. 2(b) | Fig. 2(c) | Fig. 2(d) | - | - | - |
| Space group | No. 230 | No. 088 | No. 120 | No. 220 | - | - | - |
| | No. 206 | No. 015 | No. 045 | No. 199 | - | - | - |
| $\mathcal{PT}$ | $\sqrt{}$ | $\sqrt{}$ | $\times$ | $\times$ | $\times$ | $\times$ | $\sqrt{}$ |
| $G_i\mathcal{T}$ | $\sqrt{}$ | $\sqrt{}$ | $\times$ | $\times$ | $\times$ | $\times$ | $\times$ |
| $G_{j,k}\mathcal{T}$ | $\sqrt{}$ | $\times$ | $\sqrt{}$ | $\times$ | $\times$ | $\times$ | $\times$ |
| $C_3$ | $\sqrt{}$ | $\times$ | $\times$ | $\sqrt{}$ | $\times$ | $\times$ | $\times$ |
| $l_0, l_1, l_2$ | $0,0,0$ | $l_0,0,0$ | $0,l_1,0$ | $0,0,l_2$ | $0, |l_1|<|l_2|$ | $0, |l_1|>|l_2|$ | - |
| Hamiltonian | $H(\boldsymbol{k}) \sim H_D(\boldsymbol{k}) + l_0\Gamma_{13} + l_1\Gamma_{03} + l_2\Gamma_{30}, \quad (l_0, l_1, l_2 \in \Re)$ $H_D(\boldsymbol{k}) \sim (\sigma_x + w\sigma_y) \otimes (k_z\sigma_z + k_x\sigma_y - k_y\sigma_x)$ | | | | | | |

TABLE IV: Possible results of the Dirac point transition under symmetry breaking.

## IV. REALIZATION OF VARIOUS SYMMETRY BREAKINGS

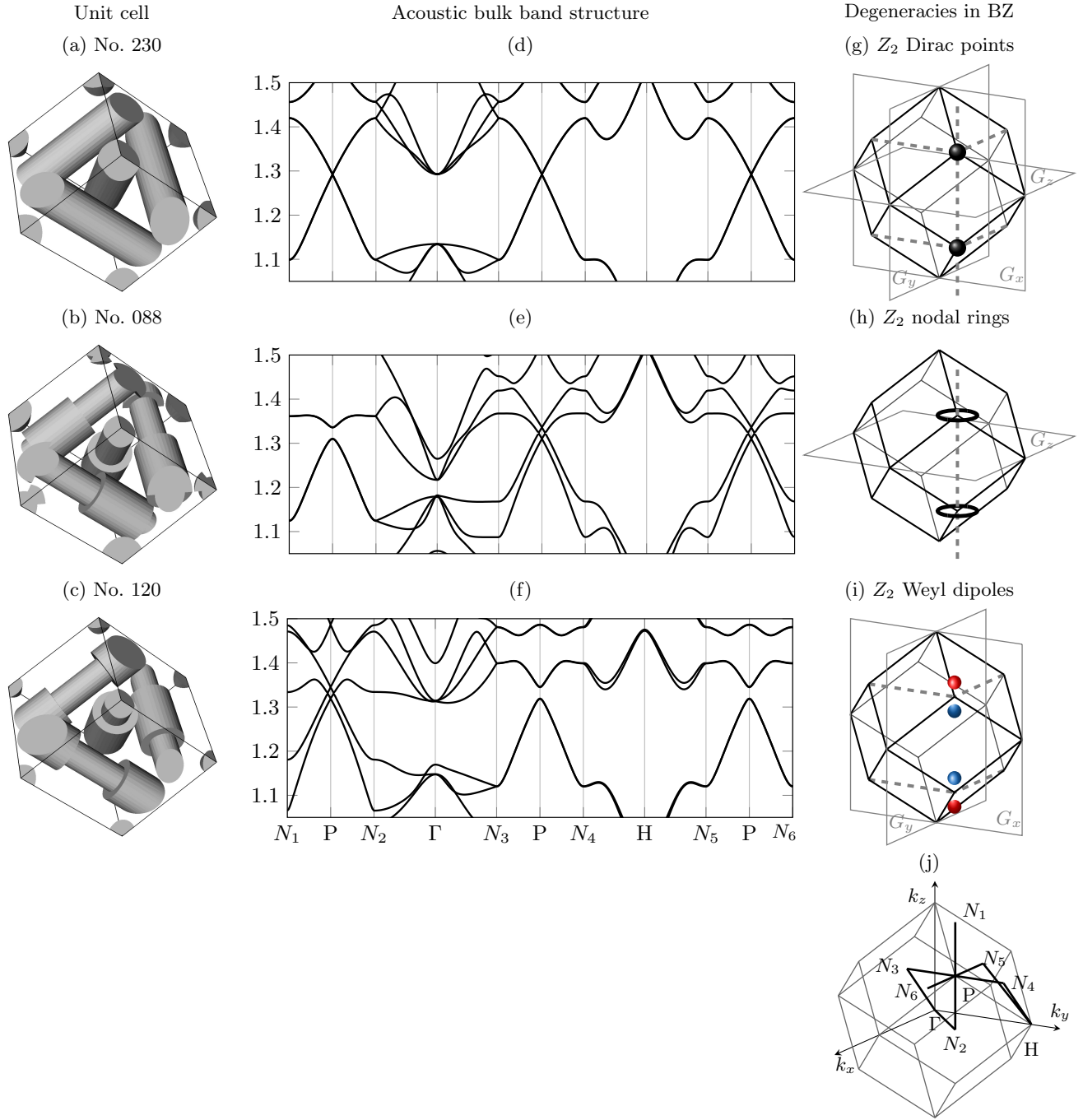

FIG. S2: Realization of the predicted symmetry breaking results. (a), (d), (g) Two Dirac points (black spheres) in the first BZ, locate at point $\pm P$. The dashed gray lines denote for the $G_i\mathcal{T}$ stabilized line degeneracies. (b), (e), (h) The Dirac points turn into $Z_2$ nodal rings (black circle). The $G_z\mathcal{T}$ stabilized degenerate lines, between lower or higher two bands, perpendicularly cross the rings and form nodal links within the four bands. (c),(f),(i) The Dirac points split into $Z_2$ Weyl dipoles (blue and red spheres for opposite chiralities). Since the two surface glides are not broken, the quad-helicoid surface states are still robust. Note that each one of $G_x\mathcal{T}$, $G_y\mathcal{T}$ is enough for keeping the $Z_2$ Weyl dipoles, but we need both to keep the quad-helicoid on 001 surface.

## V. COMPARISON OF EIGENVALUE AND LDOS CALCULATIONS

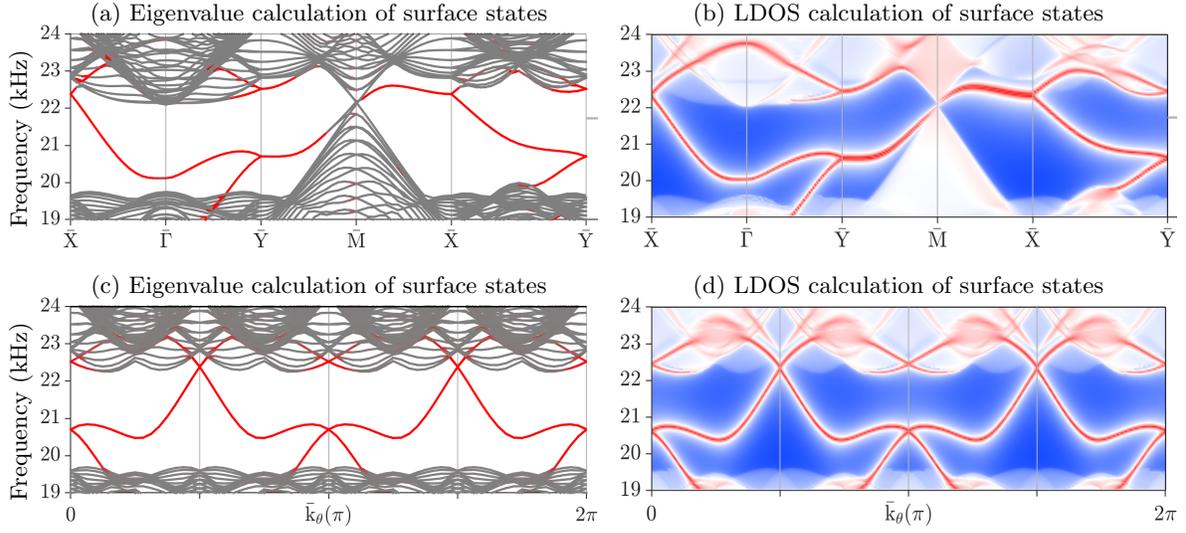

FIG. S3: Surface states in eigenvalue solutions and LDOS results. (a),(c) The surface states dispersion are obtained by solving the eigenvalue equation for a ten cubic cell stack with Floquet periodic boundary conditions in x and y directions and hard-wall boundaries on z and -z ends. The dispersion lines were colored in red if they have a large energy fraction in the first two cells on the surface. The results agree the local density of states (LDOS) results in (b),(d), respectively.

## VI. ADDITIONAL EXPERIMENTAL RESULTS

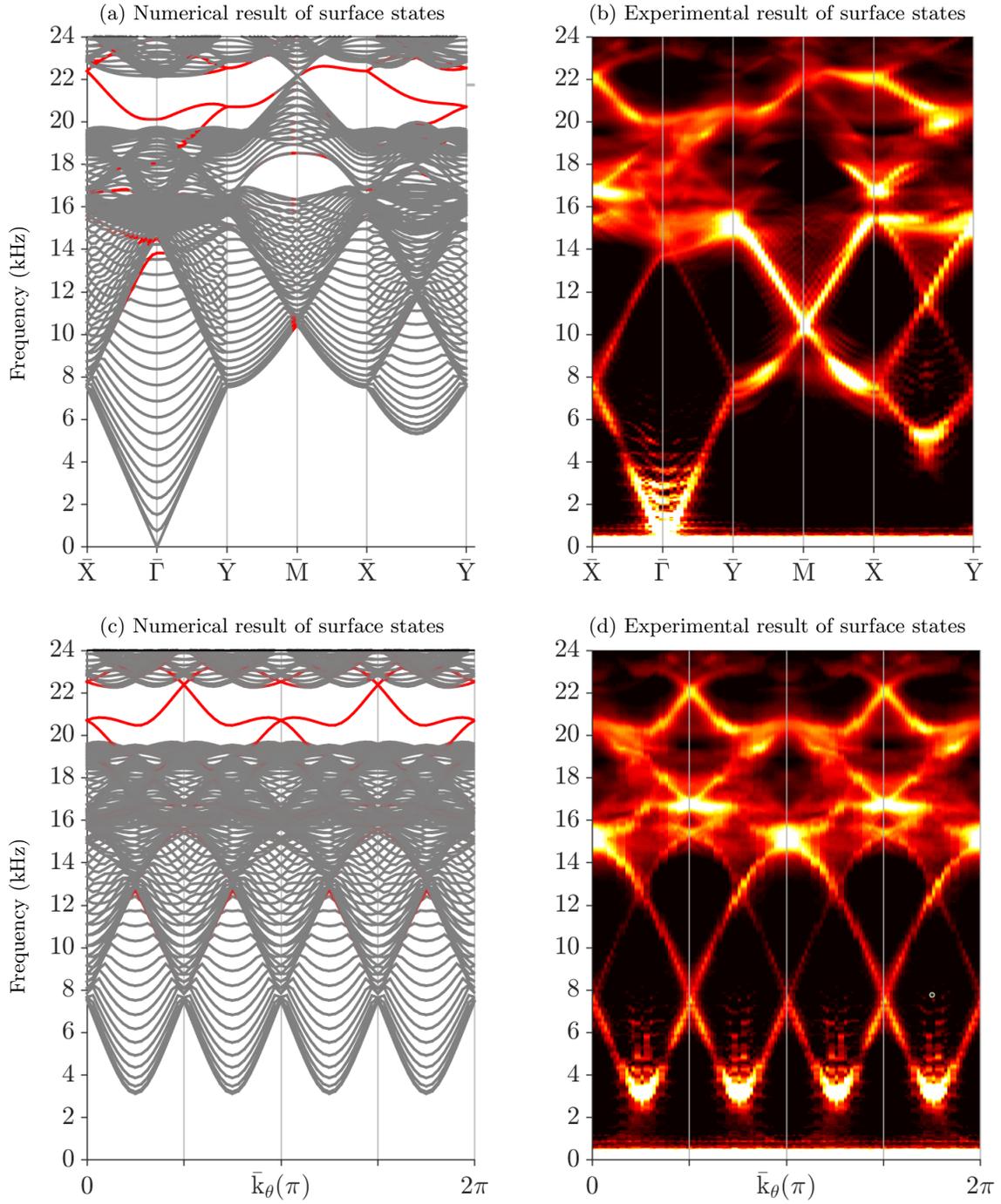

FIG. S4: Full frequency range data of the numerical and measured surface states.

## VII. MORE DETAILS ON EXPERIMENTAL MEASUREMENTS

|  | Mass density $\rho$ $(kg/m^3)$ | Longitudinal sound velocity $v_L$ $(m/s)$ | Acoustic impedance $Z = \rho \cdot v_L$ $(N \cdot s/m^3)$ |
|---|---|---|---|
| Air | 1.2 | 343 | 412 |
| Cured resin | $\sim$1140 | $\sim$2240 | $\sim$2.55e6 |

TABLE V: Material properties of air and cured resin of 3D printing. Due to the 6200 times higher acoustic impedance of resin, the air-resin interface can be treated as hard-wall boundary conditions in calculations. The pressure field inside the resin are ignored.

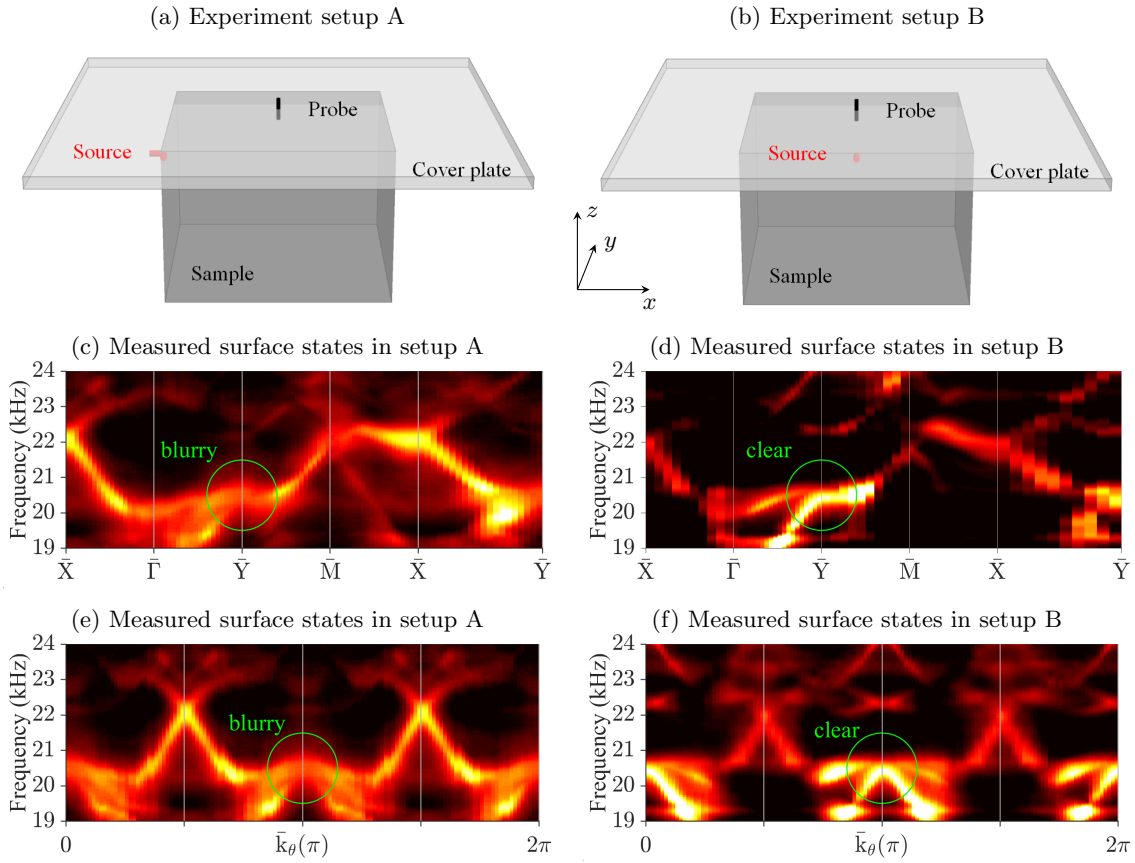

FIG. S5: Experimental data from different positions of the source excitations. (a) Setup A with the source fixed at the top corner of the sample, whose measurement results are shown in (c) and (e). (b) Setup B with the source fixed at the center of top edge of the sample, whose measurement results are shown in (d) and (f).



\end{document}